\documentclass[12pt]{article}
\usepackage{amsmath}
\usepackage{sc3conf}
\usepackage{amsfonts}
\usepackage{epsfig}

\begin{document}

\title{MSFT :\\
Moyal Star Formulation \\
of String Field Theory{\ \thanks{%
Lecture to appear in the proceedings of the Third Sakharov Conference,
Moscow, June 2002.}} }
\author{}
\maketitle

\begin{abstract}
The Moyal star formulation of string field theory is reviewed. The various
versions of the star product are compared and related to one another in a
regulated theory that resolves associativity anomalies. A summary of
computations and challenges is given.
\end{abstract}

\authors{Itzhak Bars}

\addresses{Department of Physics and Astronomy \\
University of Southern California, Los Angeles, CA 90089-0484, USA
}

\raggedbottom

\section{Introduction}

Witten's formulation of open bosonic string field theory (SFT) \cite{Witten}-%
\cite{chan} has experienced a rebirth through new physical insights and
technical advances \cite{sen}-\cite{BKM1}. Its application to certain
nonperturbative aspects of string theory, in particular D-branes, has
prompted the development of new mathematical tools to reformulate and
analyze perturbative as well as nonperturbative string physics in the
context of string field theory.

Over a year ago it was shown in \cite{witmoy} that the star product in SFT,
originally defined by Witten as a path integral that saws two strings into a
third one, is equivalent to the Moyal star product \cite{JMoyal} which is at
the heart of the familiar formulation of noncommutative geometry \cite%
{DouglasNekrasov}. We will refer to this formulation of SFT as the Moyal
Star Formulation of String Field Theory (MSFT) \cite{witmoy},\cite{anomaly}-%
\cite{chu},\cite{BKM1}.

In \cite{witmoy} it was shown that the Moyal star in MSFT is defined in the
phase space of \textit{only even string modes }$\left( x_{e}^{\mu
},p_{e}^{\mu }\right) ,$ $e=2,4,6,\cdots ,$\textit{\ } \textit{for each mode
independently, }and furthermore, that the product is local at the midpoint
of the string $\bar{x}^{\mu }$. The midpoint details were fully clarified in
\cite{moyal}. Thus, in MSFT a string field is denoted by $A\left( \bar{x}%
,\xi \right) ,$ where $\xi _{i}=\left( x_{2},x_{4},x_{6},\cdots
,p_{2},p_{4},p_{6},\cdots \right) $ is the noncommutative space formed by
the direct product of Moyal planes. The Moyal star product $\left( A\left(
\bar{x}\right) \star B\left( \bar{x}\right) \right) \left( \xi \right) $ is
defined for each pair $\left( x_{e}^{\mu },p_{e}^{\mu }\right) $
independently for each $e,$ at fixed $\bar{x}^{\mu }.$

In the discrete basis labelled by $e$ a cutoff method was introduced to
regulate associativity anomalies \cite{anomaly} and provide a precise
definition of MSFT with a regulator \cite{anomaly}\cite{moyal}.
Subsequently, a Moyal star related to the original one in \cite{witmoy} was
discussed in continuous bases \cite{DLMZ}-\cite{chu}. With proper care of
anomalies and regulators, the discrete and continuous bases are completely
equivalent as discussed in \cite{moyal} and in the next section. So, MSFT
may be pursued in different bases and it may be sometimes beneficial to
change bases for the efficiency or clarity of a computation. So far explicit
computations have been carried out mainly in the original discrete basis
which provides a natural and consistent regulator that is necessary for
careful computation.

The MSFT star product is similar to, but is different than, deformation
quantization \cite{deformation} of string theory, since in deformation
quantization all phase space degrees of freedom enter in the Moyal star,
while in MSFT the degrees of freedom involved in the Moyal star are \textit{%
half} of all phase space of the \textit{excited} modes of the open string
(also the midpoint is not part of this space).

The advantage of MSFT is its structural and computational simplicity.
Computations in MSFT are based only on the use of the Moyal star product.
The new star provides an alternative to the oscillator tool or the conformal
field theory tool as a method of computation.

To perform explicit computations, a monoid algebra (almost a group) was
introduced in \cite{witmoy}\cite{moyal} as an effective tool. The structure
of the monoid algebra turns out to be sufficient to evaluate the star
products and traces in most computations of physical interest.

In \cite{moyal} it was shown that interaction vertices in MSFT are in full
agreement with other versions of SFT, in particular the oscillator version
in \cite{Witten}. Furthermore new results involving interaction vertices for
any number of perturbative or nonperturbative string states (projectors that
describe D-branes) were obtained through MSFT for the first time. The
computation of Feynman graphs for interacting strings, with perturbative or
nonperturbative external states, has also been developed more recently using
MSFT techniques \cite{BKM1}\footnote{%
The computation of perturbative Feynman graphs is pursued in the oscillotor
formulation in \cite{taylor}.\label{feynmosc}}. As an aside it was shown
that the propagator in MSFT is also in full agreement with other approaches
to string theory. Therefore MSFT, with a cutoff, reproduces the correct
vertices and propagators of perturbative string theory while providing a
convenient formalism for pursuing nonperturbative aspects.

The progress in MSFT during the past year will be outlined, and some open
problems will be mentioned at the end.

\section{Action}

The action of open string field theory is given by \cite{Witten}%
\begin{equation}
S=\int \left( d\bar{x}^{\mu }\right) ~e^{-i\frac{3}{2}\bar{x}_{27}}~Tr\left(
\frac{1}{2}A\star \left( \mathcal{Q}A\right) +\frac{g}{3}A\star A\star
A\right) .
\end{equation}%
\newline
Here there are three ingredients: an associative star product $\star $, the
\textquotedblleft trace" Tr, and the kinetic operator $\mathcal{Q}$. This
action is similar to the Chern-Simons action. In this analogy $A\left( \bar{x%
}\right) $ is similar to a 1-form which is a matrix and is a function of the
string midpoint $\bar{x}$, $\mathcal{Q}$ behaves like an exterior
derivative, $\star $ is like a matrix product, and Tr is like a matrix
trace. Just like the Chern-Simons theory, the action is gauge invariant
under
\begin{equation}
\delta A=\mathcal{Q}\Lambda +\Lambda \star A-A\star \Lambda .
\label{gaugesymm}
\end{equation}

The specific form of $\mathcal{Q}$ defines the kinetic term, which
corresponds to a description of a vacuum in the absence of the cubic
interaction. If the cubic term is treated perturbatively, each choice of $%
\mathcal{Q}$ defines a different perturbative expansion. The perturbative
vacuum around any conformal theory background is given by the standard BRST
operator constructed from ghosts and the Virasoro operators in that
background (e.g. the flat background). However, one may expand the theory
around any nonperturbative classical solution of the action, including
solutions that correspond to D-brane backgrounds. In that case $\mathcal{Q}$
contains information about the D-brane solution while satisfying the desired
properties, namely that it acts like an exterior derivative. The vacuum
string field theory (VSFT) proposal \cite{RSZ} corresponds to the closed
string tachyon vacuum, and there is strong evidence that this is represented
by a $\mathcal{Q}$ that depends only the midpoint coordinates.

There are several realizations of $\star ,$Tr,$\mathcal{Q}$ in various
formalisms, including conformal field theory and oscillators \cite{Witten},
which have proven to be cumbersome to manipulate in computations, although
they are perfectly correct. In MSFT the star $\star $ is the standard Moyal
product in half of the phase space of the string (even modes or related, see
below) and the trace is the integral over the phase space. $\mathcal{Q}$
maintains the same structure as other formalisms but is rewritten in terms
of half the phase space, and becomes a second order differential operator in
noncommutative space. With these ingredients string field theory becomes
essentially field theory in noncommutative space, albeit with a kinetic term
that is more involved than the usual noncommutative field theory. Thus MSFT
is a much more familiar setting which is accessible to a larger audience
that has been studying noncommutative geometry, deformation or geometric
quantization, and the associated field theories. Techniques employed in
these other fields can now be borrowed to study string field theory. There
are however, fascinating new properties not seen before in noncommutative
field theory, including two timelike coordinates\footnote{%
Traditional noncommutative field theory has only coordinates with Euclidean
signature. However, in the MSFT approach we find that string field theory is
a noncommutative field theory that contains 4 timelike coordinates for each
even mode, namely $x_{e}^{\mu },p_{e}^{\mu }$ when $\mu =0,0^{\prime }$. The
label $\mu =0$ corresponds to the usual timelike coordinate. Effectively
there is a second timelike coordinate $\mu =0^{\prime }$ associated with the
bosonization of the $bc$-ghosts, $c=e^{i\phi },b=e^{-i\phi }.\;$The field $%
\phi \left( \sigma \right) $ behaves like a 27th dimension $\phi
=x^{27}=ix_{0^{\prime }}$ which is actually timelike $x_{0^{\prime }}$. The
reason is that $b,c$ are hermitian and also the string field $A$($\xi ^{
\tilde{\mu}},\xi ^{27}$) is real under the complex conjugation operation $%
A^{\ast }$($\xi ^{\tilde{\mu}},-\xi ^{27}$)$=A$($\xi ^{\tilde{\mu}},\xi
^{27} $), where $\tilde{\mu}$ represents the usual coordinates. These are
possible only if $\xi _{i}^{27}=$($\phi _{e},p_{\phi _{e}}$) are
antihermitian, or $\phi \left( \sigma \right) =ix_{0^{\prime }}\left( \sigma
\right) $ with the timelike $x_{0^{\prime }}\left( \sigma \right) $
hermitian like the other string coordinates. In conformal field theory the
operator products of $\phi $ have spacelike signature, therefore $%
x_{0^{\prime }}$ has timelike signature. Thus, in string field theory, if
the bosonized ghost is treated like the other coordinates under hermitian
conjugation, the spacetime index on $\xi _{i}^{\mu }=\left( x_{e}^{\mu
},p_{e}^{\mu }\right) $ is labelled as $\mu =0,1,\cdots 25,0^{\prime },$
with two times. The (25,2) Lorentz symmetry of $\eta _{\mu \nu }$ is broken
because the center of mass of the ghost coordinate $\phi _{0}$ is
compactified$^{\ref{compact}}$, and also the kinetic operator $\mathcal{Q}$
treats the ghosts $\phi \left( \sigma \right) =ix_{0^{\prime }}\left( \sigma
\right) $ differently than the other 26 dimensions. Nevertheless, the Moyal
star among the higher modes $\left( x_{e}^{\mu },p_{e}^{\mu }\right) $,
which determines the interactions, respects the (25,2) symmetry. Thus the
extra timelike coordinate is a fact of life in the theory even in the
absence of a full (25,2) symmetry. This situation is reminiscent of Two-Time
physics field theory \cite{twoT} which has a gauge symmetry quite similar to
the one in MSFT. It is my approach to Two-Time Physics field theory that led
me to the concept of the Moyal product in string field theory. \label{ghost}}
and the correct amount of gauge symmetry to remove related ghosts (negative
norm states caused by the timelike coordinates).

In vacuum string field theory $\mathcal{Q}$ may be as simple as $\mathcal{Q}
\left( \frac{\pi }{2}\right) \sim c^{+}\left( \frac{\pi }{2}\right)
+c^{-}\left( \frac{\pi }{2}\right) $, where $c^{\pm }\left( \sigma \right) $
are fermionic ghosts \cite{RSZ}. In the bosonized ghost approach$^{\ref%
{ghost}}$ this midpoint $\mathcal{Q}\left( \frac{\pi }{2}\right) $ becomes a
simple differential operator \cite{moyalbrst}%
\begin{equation}
\mathcal{Q}\left( \frac{\pi }{2}\right) \sim e^{i\bar{\phi}}\sin \left(
\frac{\pi }{4}-\frac{\pi }{2}\frac{-i\partial }{\partial \bar{\phi}}\right) .
\label{boseQ}
\end{equation}%
It depends only on the midpoint of the bosonized ghost $\bar{\phi}$, while $%
\frac{-i\partial }{\partial \bar{\phi}}$ is the ghost number operator. For
fields of definite ghost number we have\footnote{%
To produce the effect of the $bc$ ghosts, the zero mode of $\phi $ is
compactified on the periodic interval 0$\leq \phi _{0}\leq 4\pi $ so that
the ghost number $p_{0}^{\phi }=\frac{-i\partial }{\partial \phi _{0}}=\frac{%
-i\partial }{\partial \bar{\phi}}$ has eigenvalues $p_{0}\rightarrow n-1/2$
with $n\in
\mathbb{Z}
.$\label{compact}} $A\left( \bar{\phi}\right) \sim e^{i\left( n-\frac{1}{2}%
\right) \bar{\phi}}A_{n-\frac{1}{2}},$ $n\in
\mathbb{Z}
,$ where $A_{n-\frac{1}{2}}$ is independent of the midpoint $\bar{\phi}.$ On
this space it is easy to see that this $\mathcal{Q}$ squares to zero $%
\mathcal{Q}^{2}=0$, and it has trivial cohomology - i.e. any field that
satisfies $\mathcal{Q}A=0$ ($n$=odd) is of the form $A\sim \mathcal{Q}%
A^{\prime }$ ($n^{\prime }$=even). Therefore in this vacuum there are no
open strings since no physical solutions exist to the linearized equations
of motion $\mathcal{Q}A=0$. In this formalism, the physical string field in
the action has ghost number -1/2 ($n=0$), i.e. $A\left( \bar{\phi}\right)
\sim e^{-\frac{i}{2}\bar{\phi}}A_{-\frac{1}{2}},$ while the star product is
defined with an extra midpoint phase $\exp \left( -i\frac{3}{2}\bar{\phi}%
\right) .$ If one follows the dependence on the midpoint $\bar{\phi}$ it is
easy to see that $\bar{\phi}$ can be completely removed from the theory and $%
\mathcal{Q}$ is then replaced by a constant when applied on the physical
field. The constant can be absorbed into renormalizations of $A_{-\frac{1}{2}%
}$ and $g.$ Then the classical equation of motion becomes the projector
equation $A_{-\frac{1}{2}}\star A_{-\frac{1}{2}}=A_{-\frac{1}{2}}$ which has
many nonperturbative solutions that include D-branes. Solutions to the
projector equation are easily obtained when $\star $ is the Moyal product in
MSFT, and a general explicit class that includes the sliver, butterfly etc.
is given in \cite{moyal}. The bosonized $\mathcal{Q}\left( \frac{\pi }{2}%
\right) $ in Eq.(\ref{boseQ}) gives the simplest version of the VSFT
proposal. But the VSFT proposal is still under investigation and it appears
that the midpoint structure may receive some more clarification \cite{BKM1}.

When $Q$ corresponds to the perturbative vacuum, the action in the Siegel
gauge simplifies to%
\begin{equation}
S=\int d\bar{x}^{\mu }~Tr\left( \frac{1}{2}A\ast \left( \left(
L_{0}-1\right) A\right) -\frac{g}{3}A\ast A\ast A\right)  \label{pertS}
\end{equation}%
where $L_{0}$ is the Virasoro operator in any exact conformal theory
background. In the flat background it is a sufficiently simple differential
operator (see below) that is easily manageable in MSFT as shown in Feynman
graph computations in \cite{BKM1}. In backgrounds that are closely related
to flat backgrounds (such as tori, orbifolds, etc.) we expect that the
formalism is also reasonably manageable. The form of Eq.(\ref{pertS}) makes
it evident that MSFT is a noncommutative field theory of the usual kind, but
in a direct product space of Moyal planes\footnote{%
Should one worry that in this formulation of string field theory there are
an infinite number of timelike derivatives generated by the Moyal star? The
first thing to note is that \textquotedblleft \textit{time}" $t$ in string
theory refers to the center of mass timelike coordinate $x_{0}^{0}$, and
that in MSFT $\partial _{t}$ is rewritten as a derivative with respect to
the \textit{midpoint only} $\partial _{t}=\partial _{\bar{x}^{0}}.$ With
respect to the midpoint there are a maximum of two time derivatives in $%
L_{0} $, as in usual field theory (see Eq.(\ref{Lo})). The timelike
derivatives with respect to the higher modes $\left( x_{e}^{\mu },p_{e}^{\mu
}\right) ,$ with $\mu =0,0^{\prime },$ in the Moyal product are not the
\textquotedblleft \textit{time}" derivatives, so these do not pose a
problem. Therefore, MSFT does not suffer from multiple \textquotedblleft
\textit{time}" derivatives with respect to the actual \textquotedblleft
\textit{time}" parameter. One should worry about negative norm states
associated with timelike directions of the higher modes, but string theory
is expected to be free of negative norm states thanks to the gauge symmetry
of Eq.(\ref{gaugesymm}) combined with the on-shell BRST condition that is
equivalent to the equations of motion (including interaction).}.

\section{Equivalent star products $\star _{e},\star _{o},\star _{T},\star
_{t},\star _{\protect\kappa },\star _{\protect\sigma }$}

We turn to the star product. By now it has become clear that the MSFT star
product in \cite{witmoy}\cite{moyal} can be rewritten in many equivalent
bases. I will describe a few of these below.

Recall that the open string is parametrized in terms of string modes
\begin{equation}
x^{\mu }(\sigma )=x_{0}^{\mu }+\sqrt{2}\sum_{n\geq 1}^{\infty }x_{n}^{\mu
}\cos (n\sigma ).  \label{mode-expansion}
\end{equation}%
It will be useful to separate the modes into even $x_{e}^{\mu }$, $%
e=2,4,6,\cdots ,$ and odd $x_{o}^{\mu },$ $o=1,3,5,\cdots .$ The string
field in position space is denoted by $\psi \left( x_{0},x_{e},x_{o}\right)
. $ Witten's star product corresponds to sawing two strings into a third one
in position space - this is a complicated expression which has been
difficult to manipulate in computations.

The essential steps to the Moyal formulation of string field theory were
given in \cite{witmoy}. It starts with working in the Fourier basis $A\left(
\bar{x},x_{e},p_{o}\right) $ $\leftrightarrow \psi \left(
x_{0},x_{e},x_{o}\right) $ reached by transforming the $x_{o}$ variable to
Fourier space using the kernel $\exp \left( -\frac{2i}{\theta }%
x_{o}p_{o}\right) $. The parameter $\theta $ absorbs units and can be mapped
to 1, if desired, with an appropriate choice of units of $p_{o}$. We also
rewrite the center of mass mode in terms of the midpoint $\bar{x}=x\left(
\frac{\pi }{2}\right) $ and $x_{e}$ as
\begin{equation}
x_{0}=\bar{x}+w_{e}x_{e}.
\end{equation}

Starting with the basis $\left( x_{e},p_{o}\right) $ we will be interested
in other bases $\left( x_{e},p_{e}\right) $, $\left( x_{o},p_{o}\right) $, $%
\left( \tilde{x}_{e},\tilde{p}_{o}\right) $, $\left( x\left( \sigma \right)
,p\left( \sigma \right) \right) $, $\left( x\left( \kappa \right) ,p\left(
\kappa \right) \right) $ that are transformed into each other. The matrices $%
T_{eo},R_{oe},w_{e},v_{o},$ given by
\begin{equation}
T_{eo}=\frac{4o~i^{o-e+1}}{\pi \left( o^{2}-e^{2}\right) },\;R_{oe}=\frac{%
4e~i^{o-e+1}}{\pi o\left( o^{2}-e^{2}\right) },\;w_{e}=\sqrt{2}%
i^{-e+2},\;v_{o}=\frac{2\sqrt{2}}{\pi }\frac{i^{o-1}}{o}.  \label{TRwv}
\end{equation}%
play a special role and appear in all computations of physical quantities.
These emerge from the properties of the even and odd trigonometric functions
in Eq.(\ref{mode-expansion}). It is useful to note that they satisfy the
matrix relations (a bar means transpose)%
\begin{equation}
R=\left( \kappa _{o}\right) ^{-2}\bar{T}\left( \kappa _{e}\right) ^{2},\quad
R=\bar{T}+v\bar{w},\quad v=\bar{T}w,\quad w=\bar{R}v.  \label{define}
\end{equation}%
where $\kappa _{e},\kappa _{o}$ are diagonal matrices that represent the
string oscillator frequencies. For an infinite number of modes the
frequencies are given by $\kappa _{e}=e,$ $\kappa _{o}=o,$ but in a cutoff
theory described below the diagonal entries $\left( \kappa _{e},\kappa
_{o}\right) $ can be more general functions of $\left( e,o\right) $ while
Eqs.(\ref{define}) continue to hold.

A cutoff is needed in all formulations of SFT to resolve associativity
anomalies \cite{anomaly}\footnote{%
As seen from Eq.(\ref{TRwv}) for large $N$ we get $\bar{w}w\rightarrow
2N\rightarrow \infty $. This behavior causes associativity anomalies in
multiple matrix products. As an illustration consider the matrix product $%
RTv:$ if we use $RT=1$ and $Tv=0$ which are valid at large $N,$ we get $%
\left( RT\right) v=1\times v=v,$ versus $R\left( Tv\right) =R\times 0=0.$
Such associativity anomalies are avoided by computing all matrix products at
finite $N,$ and taking the limit only at the end. Note from Eq.(\ref%
{relations}) that the factor $\left( 1+\bar{w}w\right) \rightarrow \infty $
appears in the denominator of $Tv$ but in the numerartor of $Rw.$ Such
factors cancel each other in multiple products and resolve the associativity
anomalies.\label{anom}}. The cutoff consists of working with a finite number
of string modes $n=1,2,\cdots ,2N$ that have oscillator frequencies $\kappa
_{n}$, and introducing finite $N\times N$ matrices $%
T_{eo},R_{oe},w_{e},v_{o} $ as functions of the diagonal matrix $\kappa
=diag\left( \kappa _{e},\kappa _{o}\right) $. For $\kappa _{n}=n$ and $%
N=\infty ,$ the finite matrices $T,R,w,v$ reduce to the expressions in Eq.(%
\ref{TRwv}). The finite matrices $T,R,w,v$ are introduced as the solutions
of Eq.(\ref{define}) which are taken as the defining relations, since these
are also satisfied by the infinite matrices. These equations were uniquely
solved at finite $N$ in terms of arbitrary $\kappa _{n}$ \cite{moyal}\cite%
{BKM1}%
\begin{eqnarray}
&&T_{eo}=\frac{w_{e}v_{o}\kappa _{o}^{2}}{\kappa _{e}^{2}-\kappa _{o}^{2}}%
,\quad R_{oe}=\frac{w_{e}v_{o}\kappa _{e}^{2}}{\kappa _{e}^{2}-\kappa
_{o}^{2}}, \\
&&w_{e}={i^{2-e}}\frac{\prod_{o^{\prime }}\left\vert \kappa _{e}^{2}/\kappa
_{o^{\prime }}^{2}-1\right\vert ^{\frac{1}{2}}}{\prod_{e^{\prime }\neq
e}\left\vert \kappa _{e}^{2}/\kappa _{e^{\prime }}^{2}-1\right\vert ^{\frac{1%
}{2}}},\quad v_{o}={i^{o-1}}\frac{\prod_{e^{\prime }}\left\vert 1-\kappa
_{o}^{2}/\kappa _{e^{\prime }}^{2}\right\vert ^{\frac{1}{2}}}{%
\prod_{o^{\prime }\neq o}\left\vert 1-\kappa _{o}^{2}/\kappa _{o^{\prime
}}^{2}\right\vert ^{\frac{1}{2}}}.
\end{eqnarray}%
Although the finite matrices are given quite explicitly, most computations
are done by using simple matrix relations among $T,R,w,v$ without the need
for their explicit form. The following matrix relations are derived from Eq.(%
\ref{define}) and therefore they apply at finite as well as infinite $N$ for
all choices of $\kappa _{n}$ as a function of $n$
\begin{eqnarray}
&&TR=1_{e},\quad RT=1_{o},\quad \bar{R}R=1_{e}+w\bar{w},\quad \bar{T}%
T=1_{o}-v\bar{v},  \notag \\
&&T\bar{T}=1_{e}-\frac{w\bar{w}}{1+\bar{w}w},\quad Tv=\frac{w}{1+\bar{w}w}%
,\quad \bar{v}v=\frac{\bar{w}w}{1+\bar{w}w},  \label{relations} \\
&&Rw=v(1+\bar{w}w),\quad R\bar{R}=1_{o}+v\bar{v}\left( 1+\bar{w}w\right) .
\notag
\end{eqnarray}%
It is important to emphasize that in this formalism computing with arbitrary
frequencies $\kappa _{n}$ and finite number of modes $2N,$ is as easy as
working directly in the limit. Thus, in the following the finite matrices
with arbitrary $\kappa _{n},N$ will be assumed unless otherwise specified.

\paragraph{Even base $\star _{e}:$}

In \cite{witmoy} it was shown that it is possible to disentangle Witten's
star into independent Moyal stars for each mode $e$ by defining $p_{e}$
through the equation $p_{o}=p_{e}T_{eo}.$ Then the string field $A\left(
\bar{x},x_{e},p_{o}\right) $ is rewritten in the even phase space as $%
A\left( \bar{x},x_{e},p_{e}\right) $ and the Moyal star is diagonalized $%
A\star _{e}B$ in the noncommutative space $\xi _{i}^{\mu }=\left( x_{e}^{\mu
},p_{e}^{\mu }\right) $%
\begin{equation}
\star _{e}=e^{\eta _{\mu \nu }\sigma ^{ij}\overleftarrow{\partial }_{i}^{\mu
}\overrightarrow{\partial }_{j}^{\nu }},\;\sigma _{ij}=i\theta \left(
\begin{array}{cc}
0 & 1_{e} \\
-1_{e} & 0%
\end{array}%
\right) .
\end{equation}

While the Moyal star is diagonal in this basis, the Virasoro operator $L_{0},
$ which determines the perturbative string spectrum, is not. It is given by
\cite{moyal}\cite{BKM1}%
\begin{equation}
L_{0}=\frac{1}{2}\beta _{0}^{2}-\frac{d}{2}Tr\left( \tilde{\kappa}\right) -%
\frac{1}{4}\bar{D}_{\xi }\left( M_{0}^{-1}\tilde{\kappa}\right) D_{\xi }\,+%
\bar{\xi}\left( \tilde{\kappa}M_{0}\right) \xi ,\;\;  \label{Lo}
\end{equation}%
where $\beta _{0}=-il_{s}\frac{\partial }{\partial \bar{x}}$, $D_{\xi
}=\left( \left( \frac{\partial }{\partial x_{e}}-i\frac{\beta _{0}}{l_{s}}%
w_{e}\right) ,\;\frac{\partial }{\partial p_{e}}\right) ,$ and $l_{s}=\sqrt{%
2\alpha ^{\prime }}$ is the string length. The matrices
\begin{equation}
\;\tilde{\kappa}=\left(
\begin{array}{cc}
\kappa _{e} & 0 \\
0 & T\kappa _{o}R%
\end{array}%
\right) ,\;M_{0}=\left(
\begin{array}{cc}
\frac{\kappa _{e}}{2l_{s}^{2}} & 0 \\
0 & \frac{2l_{s}^{2}}{\theta ^{2}}T\kappa _{o}^{-1}\bar{T}%
\end{array}%
\right) ,  \label{Mo}
\end{equation}%
give the block diagonal forms
\begin{equation}
M_{0}^{-1}\tilde{\kappa}=\left(
\begin{array}{cc}
2l_{s}^{2}1_{e} & 0 \\
0 & \frac{\theta ^{2}}{2l_{s}^{2}}\kappa _{e}^{2}%
\end{array}%
\right) ,\;\;\tilde{\kappa}M_{0}=\left(
\begin{array}{cc}
\frac{1}{2l_{s}^{2}}\kappa _{e}^{2} & 0 \\
0 & \frac{2l_{s}^{2}}{\theta ^{2}}T\bar{T}%
\end{array}%
\right)   \label{Mk}
\end{equation}%
after using Eqs.(\ref{define},\ref{relations}). Every term in $L_{0}$ has
diagonal matrices except for the last term. Note that $T\bar{T}$ in $\tilde{%
\kappa}M_{0}$ is almost diagonal, since $T\bar{T}=1-\frac{w\bar{w}}{1+\bar{w}%
w}$, and the second term becomes naively negligible in the large $N$ limit
since $\bar{w}w\rightarrow \infty $. A major simplification would occur if
one could neglect this term as $N\rightarrow \infty $. However, due to the
anomalies$^{\ref{anom}}$ neglecting this term gives wrong results\footnote{%
The extra term in $L_{0}$ proportional to $\left( 1+\bar{w}w\right) ^{-1}$
was missed in \cite{DLMZ} in their attempt to compare the discrete Moyal $%
\star _{e}$ of \cite{witmoy} to the continuous Moyal $\star _{\kappa }$
directly at $N=\infty ,$ and erroneously concluded that there was a
discrepancy in the string spectrum. In fact, this term is not negligible, it
contibutes through the anomaly described in footnote (\ref{anom}) to produce
the correct spectrum. \label{wrong}}. The lesson is that the large $N$ limit
should not be taken in the Lagrangian and should wait until the end of
computations, as seen in explicit examples in \cite{BKM1}.

Most of the computations in MSFT were carried out \cite{witmoy}\cite{anomaly}%
\cite{moyal}\cite{BKM1} in this basis, but they may also be carried out with
the same ease in any of the equivalent bases discussed in this section.

\paragraph{Odd base $\star _{o}:$}

In \cite{witmoy} the odd phase space, $\left( x_{o}^{\mu },p_{o}^{\mu
}\right) $ was also apparent. One may introduce $x_{o}$ through $%
x_{e}=T_{eo}x_{o}$ and write the same string field $A\left( \bar{x}%
,x_{e},p_{o}\right) $ in terms of the odd phase space $A\left( \bar{x}%
,x_{o},p_{o}\right) .$ Witten's star is again disentangled into independent
Moyal stars $A\star _{o}B$ for each $o$ in the noncommutative space $\xi
_{I}=\left( x_{o},p_{o}\right) $%
\begin{equation}
\star _{o}=e^{\eta _{\mu \nu }\sigma ^{ij}\overleftarrow{\partial }_{I}^{\mu
}\overrightarrow{\partial }_{J}^{\nu }},\;\sigma _{IJ}=i\theta \left(
\begin{array}{cc}
0 & 1_{o} \\
-1_{o} & 0%
\end{array}%
\right)
\end{equation}%
The relation between the odd/even variables
\begin{equation}
p_{o}=p_{e}T_{eo},\;x_{o}=R_{oe}x_{e}  \label{oddcanon}
\end{equation}%
is a canonical transformation that leaves the Moyal product invariant $\star
_{e}=\star _{o}$. Furthermore, using Eqs.(\ref{define},\ref{oddcanon}) we
can write $\bar{x}=x_{0}+v_{o}x_{o}$. This substitution must be made before
computing the star $\star _{o}.$ Note in particular that $\left[
x_{o},p_{o^{\prime }}\right] _{\star _{e}}=\left[ x_{o},p_{o^{\prime }}%
\right] _{\star _{o}}=i\theta \delta _{oo^{\prime }}.$ Thus, $\left(
x_{o}^{\mu },p_{o}^{\mu }\right) $ are regarded as functions of $\left(
x_{e}^{\mu },p_{e}^{\mu }\right) ,$ so that the even/odd phase spaces do not
star commute with each other.

The Virasoro operator $L_{0}$ is easily rewritten in the new basis by
applying the canonical transformation in Eq.(\ref{oddcanon}). The result is
similar to Eq.(\ref{Lo},\ref{Mk}), with the replacement of the odd basis,
with the transformed $M_{0}^{-1}\tilde{\kappa},\tilde{\kappa}M_{0}$
\begin{equation}
M_{0}^{-1}\tilde{\kappa}\rightarrow \left(
\begin{array}{cc}
2l_{s}^{2}\left( R\bar{R}\right) _{oo^{\prime }} & 0 \\
0 & \frac{\theta ^{2}}{2l_{s}^{2}}\kappa _{o}^{2}%
\end{array}%
\right) ,\;\tilde{\kappa}M_{0}\rightarrow \left(
\begin{array}{cc}
\frac{\kappa _{o}^{2}}{2l_{s}^{2}} & 0 \\
0 & \frac{2l_{s}^{2}}{\theta ^{2}}1_{o}%
\end{array}%
\right) ,
\end{equation}%
where we used $\left( \kappa _{o}\right) ^{2}=\bar{T}\left( \kappa
_{e}\right) ^{2}T$ which follows from Eqs.(\ref{define})). Note that $R\bar{R%
}=1_{o}+v\bar{v}\left( 1+\bar{w}w\right) $ is almost diagonal, but has the
diverging term with $\left( 1+\bar{w}w\right) $ in the numerator. By
contrast, in the even case this factor appeared in $T\bar{T}$ in the
denominator. Again, this divergence does not actually occur in computations
because it gets killed by a zero. By waiting to take the limit until the end
of the computation such anomalous terms are correctly treated, and give the
correct spectrum of $L_{0}$. One may choose either $\star _{e}$ or $\star
_{o}$ to formulate MSFT at finite $N,$ and obtain the same results at finite
or infinite $N.$

\paragraph{Mixed bases $\star _{T},\star _{t}:$}

One may also define MSFT in the original mixed even/odd basis $A\left( \bar{x%
},x_{e},p_{o}\right) .$ Then, the Virasoro operator $L_{0}$ is diagonal, but
the Moyal star is not diagonal \cite{witmoy}. It is given by
\begin{equation}
A\star _{T}B=A\left( \bar{x},x_{e},p_{o}\right) ~\exp \left\{ i\theta \eta
_{\mu \nu }T_{eo}\left( \overleftarrow{\partial }_{x_{e}}^{\mu }%
\overrightarrow{\partial }_{p_{o}}^{\nu }-\overleftarrow{\partial }%
_{p_{o}}^{\nu }\overrightarrow{\partial }_{x_{e}}^{\mu }\right) \right\}
B\left( \bar{x},x_{e},p_{o}\right) ,  \label{eomixed}
\end{equation}%
which gives $\left[ x_{e},p_{o}\right] _{\star _{e}or~\star _{o}}=\left[
x_{e},p_{o}\right] _{\star _{T}}=i\theta T_{eo}.$ In this basis the
noncommutative product $\star _{T}=\exp \left( \sigma _{T}\overleftarrow{%
\partial }\cdot \overrightarrow{\partial }\right) $ involves the
antisymmetric matrix
\begin{equation}
\sigma _{T}=i\theta \left(
\begin{array}{cc}
0 & T \\
-\bar{T} & 0%
\end{array}%
\right)  \label{thetaT}
\end{equation}%
In this basis $L_{0}$ is diagonal and takes the form of Eq.(\ref{Lo}) with
the replacement $\xi \rightarrow \left( x_{e},p_{o}\right) $ and the
matrices in Eq.(\ref{Mk}) are mapped to diagonal matrices
\begin{equation}
M_{0}^{-1}\tilde{\kappa}\rightarrow \left(
\begin{array}{cc}
2l_{s}^{2}1_{e} & 0 \\
0 & \frac{\theta ^{2}}{2l_{s}^{2}}\kappa _{o}^{2}%
\end{array}%
,\right) ,\;\;\tilde{\kappa}M_{0}\rightarrow \left(
\begin{array}{cc}
\frac{\kappa _{e}^{2}}{2l_{s}^{2}} & 0 \\
0 & \frac{2l_{s}^{2}}{\theta ^{2}}1_{o}%
\end{array}%
\right) .
\end{equation}

The dependence on the frequencies $\kappa _{e},\kappa _{o}$ can be partially
shifted to the star product by a rescaling of the mixed phase space
\begin{equation}
\tilde{x}_{e}=\kappa _{e}^{1/2}x_{e},\;\tilde{p}_{o}=p_{o}\kappa _{o}^{-1/2},
\label{rescaledxp}
\end{equation}%
A new star product $\star _{t}$ can be defined by using the rescaled $T$
\begin{equation}
t_{eo}=\kappa _{e}^{1/2}T_{eo}\kappa _{o}^{-1/2}=\frac{\kappa
_{e}^{1/2}w_{e}v_{o}\kappa _{o}^{3/2}}{\kappa _{e}^{2}-\kappa _{o}^{2}}
\end{equation}%
as
\begin{equation}
A\star _{t}B=A\left( \bar{x},\tilde{x}_{e},\tilde{p}_{o}\right) ~\exp
\left\{ i\theta \eta _{\mu \nu }t_{eo}\left( \overleftarrow{\partial }_{%
\tilde{x}_{e}}^{\mu }\overrightarrow{\partial }_{\tilde{p}_{o}}^{\nu }-%
\overleftarrow{\partial }_{\tilde{p}_{o}}^{\nu }\overrightarrow{\partial }_{%
\tilde{x}_{e}}^{\mu }\right) \right\} B\left( \bar{x},\tilde{x}_{e},\tilde{p}%
_{o}\right)
\end{equation}%
The rescaled phase space satisfies $\left[ \tilde{x}_{e},\tilde{p}_{o}\right]
_{\star _{e}or~\star _{o}}=\left[ \tilde{x}_{e},\tilde{p}_{o}\right] _{\star
_{t}}=i\theta t_{eo}$, and in the new basis the noncommutative product $%
\star _{t}=\exp \left( \sigma _{t}\overleftarrow{\partial }\cdot
\overrightarrow{\partial }\right) $ involves the antisymmetric matrix
\begin{equation}
\sigma _{t}=i\theta \left(
\begin{array}{cc}
0 & t \\
-\bar{t} & 0%
\end{array}%
\right)
\end{equation}%
In this basis $L_{0}$ remains diagonal and takes the form of Eq.(\ref{Lo})
with the replacement $\xi \rightarrow \left( \tilde{x}_{e},\tilde{p}%
_{o}\right) $ and
\begin{equation}
M_{0}^{-1}\tilde{\kappa}\rightarrow \left(
\begin{array}{cc}
\frac{2l_{s}^{2}}{\kappa _{e}} & 0 \\
0 & \frac{\theta ^{2}\kappa _{o}}{2l_{s}^{2}}%
\end{array}%
\right) ,\;\;\tilde{\kappa}M_{0}\rightarrow \left(
\begin{array}{cc}
\frac{\kappa _{e}}{2l_{s}^{2}} & 0 \\
0 & \frac{2l_{s}^{2}}{\theta ^{2}\kappa _{o}}%
\end{array}%
\right) .
\end{equation}

\paragraph{kappa base $\star _{\protect\kappa }:$}

Motivated by the spectroscopy of Neumann coefficients in SFT, and guided by
the Moyal product $\star _{e}$ in \cite{witmoy}, a continuous Moyal basis $%
\left( x\left( \kappa \right) ,p\left( \kappa \right) \right) $ was
suggested in \cite{DLMZ}. This was defined directly for $N\rightarrow \infty
$ and it was evident that in explicit computations a regulator would be
needed at $\kappa =0$ to resolve issues such as an unpaired zero mode $%
\left( x\left( 0\right) =0,~p\left( 0\right) \neq 0\right) $ that is closely
related to the anomalous zero mode which was discussed earlier in \cite%
{anomaly}. A convenient regulator is the one already suggested before in
\cite{anomaly}. Indeed, it was shown in \cite{moyal}, that the continuous $%
\kappa $-basis is reached from a regulated discrete $k$-basis by
diagonalizing the finite $N\times N$ matrix $t$ that appears in $\star _{t}$%
\begin{equation}
t_{eo}=\sum_{k=1}^{N}\left( V^{\left( e\right) }\right) _{ek}\tau _{k}\left(
\bar{V}^{\left( o\right) }\right) _{ko}\rightarrow \int_{0}^{\infty }d\kappa
~V_{e}\left( \kappa \right) ~\left( \tanh \frac{\pi \kappa }{4}\right)
~V_{o}\left( \kappa \right) ,  \label{diagonal}
\end{equation}%
where the matrices $V^{\left( e\right) },$ $V^{\left( o\right) }$ are
orthogonal and $\tau _{k}$ are the eigenvalues (all functions of the
arbitrary $\kappa _{n},N$). At $N=\infty $ and $\kappa _{n}=n$ the discrete $%
k$ basis becomes the continuous $\kappa $ basis, the eigenvalues become $%
\tau _{k}\rightarrow \tanh \frac{\pi \kappa }{4}$ and the matrix elements of
$V^{\left( e\right) },$ $V^{\left( o\right) }$ become continuous functions
of $\kappa $, as given in \cite{DLMZ}\cite{moyal}.

The phase space in the discrete $k$ basis $\left( x_{k},p_{k}\right) $ is
related to the previous discrete bases by
\begin{eqnarray}
x_{k} &=&\tilde{x}_{e}\left( V^{\left( e\right) }\right) _{ek}=x_{e}\kappa
_{e}^{1/2}\left( V^{\left( e\right) }\right) _{ek},\;\; \\
p_{k} &=&\frac{2}{\theta }\tilde{p}_{o}\left( V^{o}\right) _{ok}=\frac{2}{%
\theta }p_{o}\kappa _{o}^{-1/2}\left( V^{o}\right) _{ok}=\frac{2}{\theta }%
p_{e}\kappa _{e}^{-1/2}\left( V^{\left( e\right) }\right) _{ek}\tau _{k}
\label{pk}
\end{eqnarray}%
The previous star products give%
\begin{equation}
\left[ x_{k},p_{k^{\prime }}\right] _{\star _{t}or~\star _{e}}=2\tau
_{k}\delta _{kk^{\prime }}\equiv \left[ x_{k},p_{k^{\prime }}\right] _{\star
_{\kappa }}  \label{stark}
\end{equation}%
where the right hand side defines $\star _{\kappa }=\exp \left( \sigma
_{\kappa }\overleftarrow{\partial }\cdot \overrightarrow{\partial }\right) $
with%
\begin{equation}
\sigma _{\kappa }=i\left(
\begin{array}{cc}
0 & 2\tau _{k} \\
-2\tau _{k} & 0%
\end{array}%
\right)
\end{equation}%
In this basis $L_{0}$ is not diagonal and takes the form of Eq.(\ref{Lo})
with the replacement $\xi \rightarrow \left( x_{k},p_{k}\right) $ and
\begin{equation}
M_{0}^{-1}\tilde{\kappa}\rightarrow \left(
\begin{array}{cc}
\bar{V}^{\left( e\right) }\frac{2l_{s}^{2}}{\kappa _{e}}V^{\left( e\right) }
& 0 \\
0 & V^{\left( o\right) }\frac{2\kappa _{o}}{l_{s}^{2}}\bar{V}^{\left(
o\right) }%
\end{array}%
\right) ,\;\;\tilde{\kappa}M_{0}\rightarrow \left(
\begin{array}{cc}
\bar{V}^{\left( e\right) }\frac{\kappa _{e}}{2l_{s}^{2}}V^{\left( e\right) }
& 0 \\
0 & V^{\left( o\right) }\frac{l_{s}^{2}}{2\kappa _{o}}\bar{V}^{\left(
o\right) }%
\end{array}%
\right) .
\end{equation}

At $N=\infty $ this discrete star product becomes the continuous star
product of \cite{DLMZ}. The continuous version of Eq.(\ref{stark}) is%
\begin{equation}
\left[ x\left( \kappa \right) ,p\left( \kappa ^{\prime }\right) \right]
_{\star _{\kappa }}=2\tanh \frac{\pi \kappa }{4}\delta \left( \kappa -\kappa
^{\prime }\right) .
\end{equation}%
This shows an anomalous behavior at $\kappa =0$ since the right hand side
vanishes $\tanh \frac{\pi \kappa }{4}\rightarrow 0$. A study of the
functions $V_{e}\left( \kappa \right) ,V_{o}\left( \kappa \right) $ show
that for small $\kappa $ we get $x\left( \kappa \right) \sim \kappa
\rightarrow 0$, but $p\left( \kappa \right) \rightarrow p\left( 0\right) $
tends to the zeroth power of $\kappa ,$ indicating the presence of an
anomalous zero mode. More precisely, by noting that $V_{o}\left( 0\right) =%
\frac{\sqrt{\pi o}}{4}v_{o}$ where $v_{o}$ is given in Eq.(\ref{TRwv}), and
using Eq.(\ref{pk}) in the $N\rightarrow \infty $ limit, we obtain the
relation between the anomalous mode in the continuous basis and the one
discussed in \cite{anomaly} which was $\tilde{p}^{\mu
}=\sum_{o>0}v_{o}p_{o}^{\mu }$, thus
\begin{equation}
\frac{2\theta }{\sqrt{\pi }}p^{\mu }\left( 0\right)
=\sum_{o>0}v_{o}p_{o}^{\mu }=\sum_{o,e,e^{\prime }>0}\bar{w}_{e}T_{eo}\bar{T}%
_{oe^{\prime }}p_{e\prime }^{\mu }=\frac{1}{1+\bar{w}w}%
\sum_{e>0}w_{e}p_{e}^{\mu }=\tilde{p}^{\mu }.
\end{equation}%
As we emphasized several times before it is dangerous to set $\left( 1+\bar{w%
}w\right) \rightarrow \infty $ before completing a computation, so $\frac{%
2\theta }{\sqrt{\pi }}p\left( 0\right) =$ $\tilde{p}$ does not necessarily
vanish. In fact, it is related to the structure $\frac{1}{1+\bar{w}w}\left(
\sum_{e>0}w_{e}p_{e}\right) ^{2}$ which is the anomalous part of $L_{0}$
that emerges from $T\bar{T}=1-\frac{w\bar{w}}{1+\bar{w}w}$ in Eq.(\ref{Mk}).
This piece in $L_{0}$ plays a nontrivial role in producing the correct
perturbative string spectrum and the correct propagator \cite{moyal}\cite%
{BKM1}. Thus $p\left( 0\right) $ is precisely the anomalous mode $\tilde{p}$
which needs to be treated just as carefully in computations in any basis.

In particular, note that the center of mass coordinate $x_{0}=\bar{x}%
+w_{e}x_{e}$ does not commute with the anomalous momentum mode $\tilde{p}$
since
\begin{equation}
\left[ x_{0},\tilde{p}\right] _{\star _{e}}=\sum_{o,e>0}\left[
w_{e}x_{e},v_{o}p_{o}\right] _{\star _{e}}=i\theta
\sum_{o,e>0}w_{e}T_{eo}v_{o}=i\theta \bar{v}v=i\theta \frac{\bar{w}w}{1+\bar{%
w}w}\rightarrow i\theta .  \label{anomcomm}
\end{equation}%
If the $N\rightarrow \infty $ limit is taken at intermediate steps (as in
footnote (\ref{anom})), ambiguous results follow depending on whether the $e$
or the $o$ sum is done first - doing the $o$ sum first would give $0\,$%
instead of Eq.(\ref{anomcomm}). So if one is working directly in the $%
N=\infty $ limit, should one give the priority to defining the center of
mass $x_{0}$ ($e$ first) or the anomalous mode $\tilde{p}$ ($o$ first)?
Evidently both are important and the question needs a careful answer. We see
that $p\left( 0\right) $ is clearly anomalous and not a constant after all,
since when carefully handled it does translate the center of mass (and has
other similar anomalous effects). As such, it participates in gauge
transformations that are related to general coordinate transformations, an
indication that closed strings relate to the anomalous mode $p\left(
0\right) \sim \tilde{p}$, as emphasized in \cite{anomaly}.

All this shows that one must be careful when computing in the continuous
basis. The regulated discrete version of the continuous basis given in Eq.(%
\ref{stark}) and in \cite{moyal} provides a careful approach. It is evident
that with proper care the continuous Moyal product $\star _{\kappa }$ is
fully equivalent to the other discrete Moyal products $\star _{e},\star
_{o},\star _{t}.$

\paragraph{Sigma base $\star _{\protect\sigma }:$}

The $\sigma $-base is directly related to the worldsheet parameter $\sigma .$
The same string field $A\left( \bar{x},x_{e},p_{o}\right) $ or $A\left( \bar{%
x},x_{o},p_{o}\right) $ is expressed in continuous phase space $A\left( \bar{%
x},x\left( \sigma \right) ,p\left( \sigma \right) \right) $ by using the
orthogonal transformation
\begin{equation}
x\left( \sigma \right) =\sqrt{2}\sum_{o>0}x_{o}\cos \kappa _{o}\sigma
,\;p\left( \sigma \right) =\frac{2\sqrt{2}}{\pi }\sum_{o>0}p_{o}\cos \kappa
_{o}\sigma ,\;0\leq \sigma \leq \frac{\pi }{2},  \label{sigcanon}
\end{equation}%
For $N=\infty $ and $\kappa _{o}=o,$ the odd cosines $\cos o\sigma $ form a
complete set of functions in the range $0\leq \sigma \leq \frac{\pi }{2}$
with Neumann boundary conditions at $\sigma =0$ and Dirichlet boundary
conditions at $\sigma =\frac{\pi }{2}.$ Note that, since $\sigma $ stops at
the midpoint, the phase space $\left( x\left( \sigma \right) ,p\left( \sigma
\right) \right) $ is half of the full string phase space, and excludes the
midpoint. Therefore under either the even $\star _{e}$ or odd $\star _{o}$
they produce a continuous delta function
\begin{eqnarray}
\left[ x\left( \sigma \right) ,p\left( \sigma ^{\prime }\right) \right]
_{\star _{e}~or~\star _{o}} &=&\left[ x\left( \sigma \right) ,p\left( \sigma
^{\prime }\right) \right] _{\star _{\sigma }}=i\theta \delta \left( \sigma
-\sigma ^{\prime }\right)  \label{sigstar} \\
\delta \left( \sigma -\sigma ^{\prime }\right) &=&\frac{4}{\pi }%
\sum_{o>0}\cos o\sigma ~\cos o\sigma ^{\prime },\;0\leq \sigma \leq \frac{%
\pi }{2}
\end{eqnarray}%
Therefore, Eq.(\ref{sigcanon}) is a canonical transformation, and allows us
to define the continuous $\sigma $ Moyal basis with $\star _{\sigma }$
defined by the right hand side of Eq.(\ref{sigstar}).

A discrete version of the sigma basis can also be defined by taking $N$
discrete points $\sigma _{k},$ $k=1,\cdots ,N$ in the interval $0\leq \sigma
_{k}\leq \frac{\pi }{2}$ and correspondingly choosing $\kappa _{e},\kappa
_{o}$ so that $\cos \left( \kappa _{o}\sigma _{k}\right) $ is an orthogonal
transformation acting on an $N$ dimensional basis. A discrete basis with
such properties can be constructed with the methods of \cite{barsminic}.

The $\sigma $ basis may be convenient to discuss the tensionless string
limit \cite{chu}\footnote{%
In an attempt to apply string field theory to the pp-wave background in a
lightcone formalism \cite{chu} the continuous $\star _{\sigma }$ basis was
arrived at with a series of tentative arguments that are less transparent
than the straightforward transformation given in Eq.(\ref{sigcanon}).}. We
note that in the tensionless limit the spectrum of strings produce an
infinite number of massless high spin fields \footnote{%
Massless high spin fields have been discussed in \cite{vasil} using a Moyal
star product. The Moyal basis in \cite{vasil} may be thought of as the
twistor version of phase space $\left( x^{\mu },p^{\mu }\right) ,$ hence
there is a close relation. In string theory there are many copies of the
phase space, one for each $e,$ and therefore many more high spin fields.
Furthermore, string fields are functions of the midpoint degree of freedom $%
\bar{x}.$}. Formulating an interacting theory of massless high spin fields
has been a challenge. However, by doing perturbation theory for a small
tension (described by large $l_{s}=\sqrt{2\alpha ^{\prime }}$ expansion of
the kinetic term $L_{0}$) we seem obtain a theory of interacting massless
high spin fields, which should be physically consistent, since it is a
different expansion of a consistent string theory.

\section{Summary of computations}

In \cite{moyal} it is shown that for most computations of interest one
should consider the field configurations that contain the general parameters
$\mathcal{N},M_{ij},\lambda _{i}^{\mu },k^{\mu }$%
\begin{equation}
A_{\mathcal{N},M,\lambda ,k}\left( \bar{x},\xi \right) =\mathcal{N}\exp
\left( -\bar{\xi}M\xi -\bar{\xi}\lambda +ik\cdot \bar{x}\right) ,
\label{external}
\end{equation}%
For perturbative states $M$ is replaced by $M_{0}$ of Eq.(\ref{Mo}), but for
D-brane states $\left( M\sigma \right) ^{2}=1$. The $A_{\mathcal{N}%
,M,\lambda ,k}\left( \bar{x},\xi \right) $ are the only field configurations
that are needed to compute Feynman graphs with perturbative or
nonperturbative external states \cite{BKM1}. These fields form a closed
algebra under the star%
\begin{eqnarray}
&&\left( \mathcal{N}_{1}\exp \left( -\bar{\xi}M\xi -\bar{\xi}\lambda
+ik_{1}\cdot \bar{x}\right) \right) \star \left( \mathcal{N}_{2}\exp \left( -%
\bar{\xi}M_{2}\xi -\bar{\xi}\lambda _{2}+ik_{2}\cdot \bar{x}\right) \right)
\notag \\
&=&\left( \mathcal{N}_{12}\exp \left( -\bar{\xi}M_{12}\xi -\bar{\xi}\lambda
_{12}+i\left( k_{1}+k_{2}\right) \cdot \bar{x}\right) \right)  \label{mono12}
\end{eqnarray}%
where the structure of $\mathcal{N}_{12},\left( M_{12}\right) _{ij},\left(
\lambda _{12}\right) _{i}^{\mu }$ is given as \cite{witmoy}\cite{moyal}
(define $m_{1}=M_{1}\sigma ,\;m_{2}=M_{2}\sigma ,\;m_{12}=M_{12}\sigma $)
\begin{eqnarray}
m_{12} &=&\left( m_{1}+m_{2}m_{1}\right) \left( 1+m_{2}m_{1}\right)
^{-1}+\left( m_{2}-m_{1}m_{2}\right) \left( 1+m_{1}m_{2}\right) ^{-1},
\label{m12} \\
\lambda _{12} &=&\left( 1-m_{1}\right) \left( 1+m_{2}m_{1}\right)
^{-1}\lambda _{2}+\left( 1+m_{2}\right) \left( 1+m_{1}m_{2}\right)
^{-1}\lambda _{1},  \label{lambda12} \\
\mathcal{N}_{12} &=&\frac{\mathcal{N}_{1}\mathcal{N}_{2}}{\det \left(
1+m_{2}m_{1}\right) ^{d/2}}e^{\frac{1}{4}\left( \left( \bar{\lambda}_{1}+%
\bar{\lambda}_{2}\right) \sigma \left( m_{1}+m_{2}\right) ^{-1}\left(
\lambda _{1}+\lambda _{2}\right) -\bar{\lambda}_{12}\sigma \left(
m_{12}\right) ^{-1}\lambda _{12}\right) }\,\,.  \label{n12}
\end{eqnarray}%
The algebra (\ref{mono12}) is a monoid, which means it is associative,
closed, and includes the identity element ($A_{1,0,0,0}=$1). It is short of
being a group since some elements (in particular projectors) do not have an
inverse, although the generic element does have an inverse. The trace of a
monoid is given by%
\begin{equation}
Tr\left( A_{\mathcal{N},M,\lambda ,k}\right) =\frac{\mathcal{N}e^{ik\cdot
\bar{x}}e^{\frac{1}{4}\bar{\lambda}M^{-1}\lambda }}{\det \left( 2M\sigma
\right) ^{d/2}}.  \label{trace}
\end{equation}

This monoid structure was used as a computational tool in \cite{moyal} to
calculate explicitly the following quantities (with perturbative $M_{0},$ or
any nonperturbative $M$)

\begin{itemize}
\item Static field configurations of interest, including: \textquotedblleft
Wedge" fields that correspond to powers of a single monoid $W_{n}\left( \xi
\right) =\left( A_{\mathcal{N},M,\lambda ,0}\right) _{\star }^{n}$; Sliver
field that corresponds to the infinite power of the perturbative vacuum $\Xi
\left( \xi \right) =\left( A_{\mathcal{N}_{0},M_{0},0,0}\right) _{\star
}^{\infty }$; D-brane vacua described by projectors that satisfy $A\star
A=A, $ with Tr$A=1.$ These projectors take the general form $A_{D,\lambda
}\left( \xi \right) =\mathcal{N}\exp \left( -\bar{\xi}D\xi -\bar{\xi}\lambda
\right) $ with
\begin{equation}
\mathcal{N}=\left( \prod_{e>0}2^{d}\right) \exp (-\frac{1}{4}\bar{\lambda}%
\sigma D\sigma \lambda ),\;\;D=\left(
\begin{array}{cc}
a & ab \\
ba & ~\frac{1}{\theta ^{2}a}+bab%
\end{array}%
\right) ,  \label{projector}
\end{equation}%
for any $\lambda $ and any symmetric $a,b.$ The components of $\lambda ^{\mu
}$ parallel to the D-brane vanish $\lambda ^{\parallel }=0,$ while those
perpendicular to the D-brane are functions of the transverse components of
the midpoint $\lambda ^{_{\bot }}\left( \bar{x}_{_{\bot }}\right) \neq 0$.
The \textquotedblleft Sliver", \textquotedblleft Butterfly" etc. are special
cases of the above form.

\item All n-point interaction vertices Tr$\left( A_{1}\star \cdots \star
A_{n}\right) $ in the cutoff theory were computed for perturbative or
nonperturbative monoids, for any frequencies $\kappa _{e},\kappa _{o},N$.
The simplicity of such computations is one of the payoffs of the
reformulation provided by MSFT. For $N\rightarrow \infty $ these
computations reproduced and generalized many results that were obtained
through other methods and produced new ones that were computed for the first
time. Such explicit analytic results, especially at finite $N,$ are new, and
not obtained consistently in any other approach. At finite $N$ the MSFT
results could be used in numerical as well as analytic computations as a
more consistent method than level truncation.

\item As a test of MSFT, Neumann coefficients used in the oscillator
formulation for any number of strings were computed as a corollary of the $n$%
-point vertices mentioned above. Previous computations of these coefficients
relied on conformal field theory, which yielded expressions that were
difficult to manipulate, and only for a few strings. The MSFT computation
was done with arbitrary oscillator frequencies $\kappa _{n}$ and cutoff $N.$
The cutoff version of Neumann coefficients $N_{mn}^{rs}\left( t\right)
,N_{0n}^{rs}\left( t,w\right) ,N_{00}^{rs}\left( t,w\right) ,$ were found to
be simple analytic expressions that depend on the single $N\times N$ matrix $%
t_{eo}=\kappa _{e}^{1/2}T_{eo}\kappa _{o}^{-1/2}$ and the $N$-vector $w_{e}$%
. These explicitly satisfy the Gross-Jevicki nonlinear relations for any $%
\kappa _{n},N$. It is then evident that $T$ and $w$ are more fundamental
than the Neumann coefficients. As a corollary of this result, by
diagonalizing the matrix $t$ as in Eq.(\ref{define}) one can easily
understand at once why there is a Neumann spectroscopy for the 3-point
vertex \cite{spectroscopy} or more generally the $n$-point vertex \cite%
{moyal}. The MSFT result for $n$ strings agreed with the conformal field
theory computation whenever the latter were available.

\item The oscillator and Virasoro algebra were constructed from fields under
the star product$.$ The fields $\mathcal{L}_{n}\left( \xi \right) $ were
constructed from half the phase space $\xi $ (unlike the usual construction
that uses the full oscillator space). These are fields, not differential
operators, and they close under the star commutator to form the Virasoro
algebra in noncommutative space. The exponentiated Virasoro fields form a
subset of the monoid algebra that corresponds to a group.

\item Tachyon condensation and small fluctuations in VSFT were investigated
(for an independent approach see \cite{okawa}). In \cite{moyal} it was
concluded that at finite $N$ the tachyon equation $\Xi \star T+T\star \Xi =T
$ had only pure gauge solutions, and that at $N=\infty $ the associativity
anomaly needed to play a role for the VSFT program to succeed. In later work
\cite{BKM0} it is found that it is possible to construct the perturbative
tachyon from VSFT provided the tachyon is guessed as a configuration that
almost obeys the VSFT equations as $N\rightarrow \infty $ thanks to the
anomaly. In the notation of \cite{moyal} the tachyon fluctuation around the
sliver field $\Xi \left( \xi \right) $ is given by $T\left( \bar{x},\xi
\right) =\frac{2^{-3/2}}{\sqrt{2\beta \tilde{K}}}\int d^{d}k~n\left(
k\right) e^{-\frac{1}{4}\bar{\lambda}\left( k\right) \sigma m\lambda \left(
k\right) }\Xi \left( \xi \right) e^{-\xi \lambda \left( k\right) }e^{ik\cdot
\bar{x}},$ where $n\left( k\right) $ is the normalized tachyon wavefunction
in momentum space, while
\begin{equation}
\lambda ^{\mu }\left( k\right) =k^{\mu }\left(
\begin{array}{c}
\left( a^{1/2}w\right) _{e} \\
0%
\end{array}%
\right) \sqrt{\frac{8}{\bar{w}w}\frac{\ln \left( F\left( k^{2}\right)
/2\right) }{k^{2}}},
\end{equation}%
where $a$ is the matrix in the sliver field (as in Eq.(\ref{projector}) with
$b=0$). The tachyon equation is obeyed since $\lambda ^{\mu }\left( k\right)
\rightarrow 0$ as $\bar{w}w\rightarrow 2N\rightarrow \infty .$ But note that
$\exp \left( \frac{1}{8}\bar{\lambda}\sigma m\lambda \right) =\frac{1}{2}%
F\left( k^{2}\right) $ is finite. This is the anomaly. By taking a form
factor whose behavior near the tachyon mass shell $k^{2}\sim -1$ is
\begin{equation}
F\left( k^{2}\right) \approx 1+\beta \left( k^{2}+1\right) +O\left( \left(
k^{2}+1\right) ^{2}\right) ,\;with\;\beta =\frac{1}{2}\left( \frac{\pi ^{2}}{%
3}\right) ^{1/3},
\end{equation}%
we find the correct relation between the D25-brane tension and the tachyon
coupling
\begin{equation}
g_{T}=\frac{1}{\sqrt{\tilde{K}}}\frac{1}{\left( 2\beta \right) ^{3/2}}%
,\;T_{25}=\frac{\tilde{K}}{6},\;\rightarrow T_{25}=\frac{1}{2\pi ^{2}}\frac{1%
}{g_{T}^{2}}.
\end{equation}%
What remains to be understood in this problem is the properties of the form
factor $F\left( k^{2}\right) $ given above. But note that the midpoint
structure of VSFT may receive some correction \cite{BKM1}, and this may also
shed some light on the remaining issues in this problem.

\item More recently the framework for computations of Feynman graphs using
MSFT was presented in \cite{BKM1}$^{\ref{feynmosc}}$. Efficient analytic
methods of computation were developed based on (a) the monoid algebra in
noncommutative space and (b) the conventional Feynman rules in Fourier
space. The methods apply equally well to perturbative string states or
nonperturbative string states involving D-branes.

\item The ghost sector was mostly handled in the bosonized version. However
it can also be formulated using Moyal products with fermionic $b,c~$ghosts
\cite{erler}\cite{BKM1}. The version in \cite{BKM1} takes into account the
subtleties of the midpoint.
\end{itemize}

\section{Challenges}

Some open problems and challenges beyond the topics discussed above are the
following:

\begin{itemize}
\item Construction of BRST operator in MSFT with fermionic or bosonized
ghosts. This is straightforward for $N=\infty ,$ but a version that applies
at finite $N$ in the cutoff theory is still lacking because a substitute for
the Virasoro algebra that closes at finite $N$ remains to be found. Using
such a BRST operator nonperturbative solutions of the equations of motion
that connect the perturbative vacuum to the nonperturbative (D-brane) vacua
can be investigated more reliably in the cutoff theory.

\item Closed string field configurations and their relation to associativity
anomalies remain to be investigated. This may be understood eventually
through the Feynman graph techniques given in \cite{BKM1}.

\item Formulation of supersymmetric MSFT. Our proposal is that this may be
achieved by figuring out the Moyal-Weyl quantization of the Brink-Schwarz
superparticle. Then MSFT fields would be taken as functions of $\left(
x_{e}^{\mu },p_{e}^{\mu },\theta _{e}^{\mu }\right) $ with a supersymmetric
Moyal product among them.

\item Formulation of MSFT for strings propagating on curved backgrounds.
Here we expect that the Kontsevich star product \cite{kontsevich} plays a
role. As we have seen in the discussion of various Moyal bases, the
complications may be shifted from the star product to the propagator or vice
versa. Naively, in curved backgrounds, it appears that all the burden may be
put on a complicated $L_{0}$ while the star product could be chosen as the
simple Moyal product based on the canonical structure of string theory
(perhaps this ignores global properties of the curved space). However, an
alternative to this could be to shift the burden to a complicated Kontsevich
star product while maintaining a simpler form of $L_{0}.$ The relation
between the geometric properties of the background that determines $L_{0}$
and the Kontsevich star product is currently poorly understood.

\item A study of high spin massless fields by expanding MSFT around
tensionless strings appears to be promising.

\item SFT or MSFT look like gauge fixed from something else since they
contain ghosts. What is the higher gauge symmetry and what is the higher
symmetric form of MSFT before gauge fixing? We suspect that general
canonical transformations in $\xi $ space may provide an answer.

\item Connection to 2T-physics. In footnote (\ref{ghost}) it was already
explained that MSFT with bosonized ghosts has a noncommutative space with $%
\left( 25,2\right) $ signature. There are some similarities between MSFT and
2T-physics field theory \cite{twoT} and a deeper relation between them is
suspected. The 2T approach displays higher hidden symmetries and provides a
holographic view of the higher dimensional theory in the form of many dual
lower dimensional theories. As recently demonstrated \cite{zero} AdS$%
_{5}\times $S$^{5}$ supergravity can be viewed as a holographic picture of a
12-dimensional 2T theory with $\left( 10,2\right) $ signature. A lot more
structure of this type in M-theory may be expected, and the MSFT and 2T
techniques seem to be natural tools to discover it and formulate it.
\end{itemize}

MSFT demystifies string field theory by providing a more familiar structure
in the form of noncommutative geometry with more efficient computational
tools. There still are technical and conceptual challenges as outlined in
this lecture, but there is good reason to believe that the tools to make
progress in those areas are available or can be developed. So we hope that
string field theory will become a viable framework for defining the
nonperturbative theory as well as for performing practical computations to
relate it to low energy physics and cosmology.

\bigskip

\begin{center}
\noindent{\large \textbf{Acknowledgments}}
\end{center}

This research was in part supported by a DOE grant DE-FG03-84ER40168. Useful
communications with Y. Matsuo and I. Kishimoto are gratefully acknowledged.

\end{document}